\documentclass[10pt, conference, letterpaper]{IEEEtran}
\IEEEoverridecommandlockouts
\usepackage{cite}

\usepackage{amsmath,amssymb,amsfonts}
\usepackage{algorithmic}
\usepackage{graphicx}
\usepackage{subfigure}
\usepackage{textcomp}
\usepackage{xcolor}
\usepackage{multirow}

\def\BibTeX{{\rm B\kern-.05em{\sc i\kern-.025em b}\kern-.08em
    T\kern-.1667em\lower.7ex\hbox{E}\kern-.125emX}}

\begin{document}

\title{GCN-GAN: A Non-linear Temporal Link Prediction Model for Weighted Dynamic Networks}

\author{
\IEEEauthorblockN{
    Kai Lei\textsuperscript{\dag,\S}, 
    Meng Qin\textsuperscript{\dag},
    Bo Bai\textsuperscript{\ddag,*},
    Gong Zhang\textsuperscript{\ddag},
    Min Yang\textsuperscript{\P,*}
    }
\IEEEauthorblockA{
    \textsuperscript{\dag}ICNLAB, School of Electronics and Computer Engineering (SECE), Peking University, Shenzhen, China \\
    \textsuperscript{\S}PCL Research Center of Networks and Communications, Peng Cheng Laboratory, Shenzhen, China\\
    \textsuperscript{\ddag}Future Network Theory Lab, 2012 Labs, Huawei Technologies, Co. Ltd., Hong Kong, China\\
    \textsuperscript{\P}Shenzhen Institutes of Advanced Technology, Chinese Academy of Sciences, Shenzhen, China\\
    \textsuperscript{\dag}leik@pkusz.edu.cn,
    \textsuperscript{\dag}mengqin\_az@foxmail.com,
    \textsuperscript{\ddag}\{baibo8, nicholas.zhang\}@huawei.com,
    \textsuperscript{\P}min.yang@siat.ac.cn\\
    \textsuperscript{*}Corresponding Authors
    }
}

\maketitle

\begin{abstract}
In this paper, we generally formulate the dynamics prediction problem of various network systems (e.g., the prediction of mobility, traffic and topology) as the temporal link prediction task. Different from conventional techniques of temporal link prediction that ignore the potential non-linear characteristics and the informative link weights in the dynamic network, we introduce a novel non-linear model GCN-GAN to tackle the challenging temporal link prediction task of weighted dynamic networks. The proposed model leverages the benefits of the graph convolutional network (GCN), long short-term memory (LSTM) as well as the generative adversarial network (GAN). Thus, the dynamics, topology structure and evolutionary patterns of weighted dynamic networks can be fully exploited to improve the temporal link prediction performance. Concretely, we first utilize GCN to explore the local topological characteristics of each single snapshot and then employ LSTM to characterize the evolving features of the dynamic networks. Moreover, GAN is used to enhance the ability of the model to generate the next weighted network snapshot, which can effectively tackle the sparsity and the wide-value-range problem of edge weights in real-life dynamic networks. To verify the model's effectiveness, we conduct extensive experiments on four datasets of different network systems and application scenarios. The experimental results demonstrate that our model achieves impressive results compared to the state-of-the-art competitors.

\end{abstract}

\begin{IEEEkeywords}
Temporal Link Prediction, Weighted Dynamic Networks, Generative Adversarial Networks, Graph Convolutional Networks
\end{IEEEkeywords}

\section{Introduction}

Dynamics is a significant factor that hinders the performance of most network systems. The prediction of mobility, traffic and topology has been considered as an effective technique to cope with the problem. For instance, the dynamics of communication links in ad hoc networks makes the design of routing protocol a challenging problem, where the prediction of the dynamic topology plays an important role to achieve a more efficient and reliable communication \cite{GHOUTI2013305}. In data center networks, traffic prediction technique could be utilized to effectively schedule the highly parallel network flows while avoiding the performance degradation due to resource shortages \cite{Mozo2018Forecasting}. For cellular networks, the prediction of users' locations can help to reduce the resource consumption (e.g., bandwidth) and achieve better Quality of Services (QoS) \cite{Prajapati2018Mobility}. In a word, if the dynamics of the network system can be accurately predicted, the key resources can be effectively pre-allocated to ensure the system's high performance.


Although numerous studies have been developed to deal with the dynamics of network systems \cite{Mozo2018Forecasting, Prajapati2018Mobility, Nie2018Network, Verma2017Multi}, most of them only focus on a specific application scenario (e.g., flow prediction in data center networks), failing to be generalized to other different scenarios.

In fact, the dynamics prediction problem of various network systems can be generally formulated as the temporal link prediction task, where the system's behavior is described by an abstracted dynamic graph. For example, one can model each host in a data center as a node (entity), and the dynamic traffic between a pair of hosts can be regarded as the changed weighted link (relation) between the corresponding entity pair. Given the graph snapshots of previous time slices, the temporal link prediction task tries to construct the graph topology in the next time slice \cite{Ma2018Graph}.

Several recent techniques have been proposed to predict the temporal links in dynamic graphs from different perspectives \cite{Mozo2018Forecasting, Prajapati2018Mobility, Nie2018Network, Verma2017Multi}. Despite their effectiveness, we argue that temporal link prediction remains a challenging task for two primary reasons.  

First, to the best of our knowledge, most of the existing approaches merely consider the link prediction in unweighted networks, determining the existence and absence of a link between a certain node pair. However, the link weights are essential in real networks, which bring significant information about the network's behavior. For example, the link weights may contain some useful information about delay, flow, signal strength or distance of the network systems. Under such circumstance, the temporal link prediction technique should not only determine the existence of links but also consider the corresponding weights, which is a more challenging problem that most of the conventional methods cannot tackle. 


Second, non-linear transformations over time are commonly observed in dynamic networks since the formation process of most networks is complicated and highly non-linear \cite{Cui2017A}. However, conventional methods are almost based on typical linear models (e.g., non-negative matrix factorization (NMF) \cite{Lee1999Learning}), ignoring the potential non-linear characteristics of dynamic networks. These linear models may have limited performances for some network inference tasks, including the temporal link prediction, because the linear data representation cannot capture the different latent factors of variation behind the network. To that end, it would be highly desirable to exploit the composition of multiple non-linear transformations of networks to improve the link prediction performance.



To alleviate the aforementioned limitations, we propose a novel deep learning based model for the temporal link prediction of weighted dynamic networks. Our model combines the strengths of the deep neural networks (i.e., graph convolutional network (GCN) \cite{Kipf2016Semi} and long short-term memory (LSTM) network \cite{Gers2014Learning}) as well as generative adversarial network (GAN) \cite{Goodfellow2014Generative} to strengthen the representation learning of the network data and generate the high-quality graph snapshot in next time slice. Concretely, we first utilize the GCN to capture the characteristics of topological structure hidden in each single graph snapshot. Then, the learned network representations are fed into an LSTM network to capture the evolving patterns of the weighted dynamic network with multiple successive time slices. Moreover, GAN is applied to generate high-quality and plausible graph snapshot with adversarial training. In the adversarial process, we train a generative model $G$ to predict the weighted links in the next time slice based on the historical data sequentially. A discriminative model $D$ is also trained to distinguish the generated list of links from the real records. $G$ and $D$ are jointly optimized with a minimax two-player game, enabling the model to generate high-quality weighted links.


We summarize our main contributions as follow:
\begin{itemize}
\item We formulate the dynamics prediction of various network systems as the temporal link prediction problem and discuss the challenges for the prediction of weighted dynamic networks.
\item We employ deep neural networks (i.e., GCN and LSTM) to explore the non-linear characteristics of topological structure and evolving patterns hidden in the network.
\item To the best of our knowledge, we are the first to utilize GAN to tackle the temporal link prediction of weighted dynamic networks, by generating high-quality next weighted links based on the historical snapshots.
\item Besides the standard mean square error (MSE) metric, we introduce two additional metrics (i.e., edge-wise KL-divergence and the mismatch rate) to investigate the sparsity of the relations among entities in network systems and the wide-value-range property of edge weights.
\item To verify the effectiveness of our model, we conduct extensive experiments on four datasets of various network systems, where our model consistently outperforms other competitors for the temporal link prediction task of weighted dynamic networks.
\end{itemize}

The rest of this paper is organized as follows. In Section \uppercase \expandafter {\romannumeral2}, we briefly introduce the related work.
A formal definition of the temporal link prediction problem is given in Section \uppercase \expandafter {\romannumeral3}. Section \uppercase \expandafter {\romannumeral4} presents the proposed GCN-GAN model in details. In Section  \uppercase \expandafter {\romannumeral5}, we describe the experiments, including the performance evaluation on four datasets of network systems and a case study of the proposed model's refining effect. Section \uppercase \expandafter {\romannumeral6} concludes this paper and indicates our future work.


\section{Related Work}
The dynamics of real network systems has received considerable attention in recent years. Several techniques have been developed to tackle the performance degradation caused by the system's dynamics.
In \cite{Mozo2018Forecasting}, the authors introduced a convolutional neural network framework, which can forecast the short-term traffic load in data center networks. In \cite{Prajapati2018Mobility}, a hidden Markov model was constructed to predict users' locations in mobile cellular networks. To improve the quality of service (QoS) for users in wireless mesh backbone networks, \cite{Nie2018Network} proposed a network traffic prediction model by integrating the deep belief neural network and spatiotemporal compressive sensing method. For the traffic matrix estimation problem, a novel approach with multiple low-rank matrices was advocated in \cite{Verma2017Multi}, achieving a better performance compared to the conventional gravity model. However, most of the dynamics prediction techniques of network systems (including the above methods) only utilize the unique patterns or characteristics of a specific application scenario (e.g., data center network), lacking the significant ability to be generalized to other different scenarios.

On the other hand, the dynamics prediction problem of network systems can be generally modeled as the temporal link prediction task, and a brief overview about the task can be found in \cite{Ma2017Nonnegative} and \cite{Huang2009The}.

Conventional temporal link prediction methods are almost based on the collapsed network model \cite{Liben2007The, Sharan2008Temporal}. In the model, the network snapshots of multiple successive time slices are linearly combined to construct a single comprehensive snapshot named as the collapsed network. The characteristics of the dynamic network are extracted by conducting a certain matrix decomposition process on the collapsed snapshot. Nevertheless, such conventional models may ignore the critical information hidden in the dynamic network with multiple network snapshots resulting in limited prediction performance.

To avoid collapsing the temporal networks, authors of \cite{Dunlavy2011Temporal} represented the dynamic network as a third-order tensor, and the temporal information was explored by conducting a tensor factorization process. In \cite{Ma2018Graph}, a model based on the non-negative matrix factorization (NMF) framework \cite{Lee1999Learning} was developed, where the dynamic information of historical snapshots was incorporated by utilizing the graph regularization technique. As discussed in \cite{Kai2018Adaptive}, each network snapshot in the dynamic network could be described as a corresponding NMF component. A unified model was proposed based on the combination of multiple NMF components, where a novel adaptive parameter was introduced to consider the intrinsic correlation between single snapshot and the dynamic network. 

However, the aforementioned approaches still have limited room for the improvement of prediction accuracy, because they are almost based on the traditional linear model, ignoring the potential non-linear characteristic of the dynamic network. Although several non-linear methods based on the restricted Boltzmann machine (RBM) \cite{Li2014A} and graph embedding \cite{Hisano2018Semi} are proposed, most of them can only be applied to the prediction of unweighted networks but cannot deal with the challenging case of weighted networks.

\section{Problem Definition}
A dynamic network can be defined as a sequence of graph snapshots $G=\left\{{{G_1},{G_2},\cdots ,{G_\tau }}\right\}$, in which ${G_t}=\left( {V,{E_t},{W_t}} \right)$ is the snapshot at a certain time slice $t$ ($t\in\left\{{1,2, \cdots ,\tau }\right\}$) with a node set $V$, a edge set $E_t$ and a weight set $W_t$ (we use the subscript $\tau$ to represent current time slice). In this study, we only consider the case of undirected weighted networks with all the graph snapshots sharing the same node set $V$.

For the snapshot of time slice $t$, we use an adjacency matrix ${{\bf{A}}_t} \in {\Re ^{\left| V \right| \times \left| V \right|}}$ to describe the corresponding static topological structure. Specially, when there is an edge between node $i$ and $j$ ($(i,j)\in{E_t}$) with weight ${W_t}\left({i,j}\right)$, we let ${\left( {{{\bf{A}}_t}} \right)_{ij}} = {\left( {{{\bf{A}}_t}} \right)_{ji}}={W_t}\left({i,j}\right)$, and  ${\left({{{\bf{A}}_t}} \right)_{ij}} = {\left( {{{\bf{A}}_t}} \right)_{ji}}=0$ otherwise.

Given the adjacency matrices of previous $l$ time slices and current time slice $\left\{ {{{\bf{A}}_{\tau - l}},{{\bf{A}}_{\tau - l + 1}}, \cdots ,{{\bf{A}}_\tau }} \right\}$ (with $l+1$ network snapshots in total), the goal of the temporal link prediction task is to predict the topology of the next time slice $(\tau +1)$, which can be formally described below:
\begin{equation}
    {{\bf{\tilde A}}_{\tau + 1}} = f\left( {{{\bf{A}}_{\tau  - l}},{{\bf{A}}_{\tau - l + 1}} \cdots ,{{\bf{A}}_\tau }} \right),
\end{equation}
where $f\left( \cdot \right)$ is the model that we need to construct in this paper while ${{\bf{\tilde A}}_{\tau + 1}}$ represent the prediction result. For the convenience of discussion, we utilize the simplified notation ${{\bf{A}}_{\tau - l}^\tau }$ to represent the sequence $\left\{ {{{\bf{A}}_{\tau - l}}, \cdots ,{{\bf{A}}_\tau }} \right\}$.

\section{Methodology}
\subsection{The Model Architecture}
In this study, we introduce a novel non-linear model GCN-GAN for the temporal link prediction of weighted dynamic networks. The proposed model, depicted in Fig. 1, consists of three main components: (\romannumeral1) Graph Convolutional Network (GCN) \cite{Kipf2016Semi}, (\romannumeral2) Long Short-Term Memory (LSTM) \cite{Gers2014Learning} and (\romannumeral3) Generative Adversarial Nets (GAN) \cite{Goodfellow2014Generative}.

First, we utilize the GCN to explore the local topology characteristics of each single graph snapshot. Then, the comprehensive representations given by the GCN are fed into an LSTM network to capture the evolving patterns of the dynamic graph. Moreover, we apply the GAN to generate high-quality predicted graph snapshot with an adversarial process, where we use the GCN as well as the LSTM to construct a generative network $G$ (bottom side of Fig. 1) and introduce another full-connected discriminative network $D$ (top side of Fig. 1). In the adversarial process, $G$ is trained to predict the next snapshot based on the dynamic graph's historical topology, while $D$ is trained to distinguish the generated weighted links from the real records. By applying this minimax two-player game, the adversarial process eventually adjusts $G$ to generate plausible and high-quality prediction result.


In the rest of this section, we elaborate on the three main components of GCN-GAN in detail.

\begin{figure}[htbp]
\centerline
{\includegraphics[width=0.5\textwidth, trim = 5 5 5 5, clip]{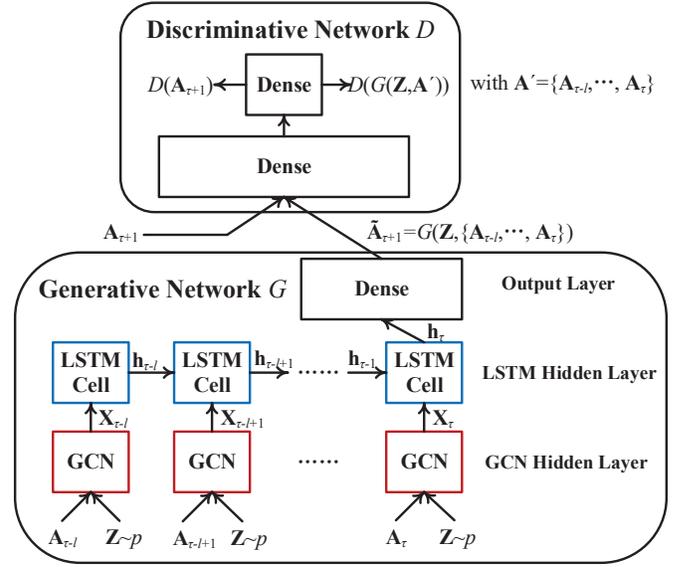}}
\caption{The architecture of the GCN-GAN temporal link prediction model with a generative network $G$ (bottom side) and a discriminative network $D$ (top side). The generative network consists of a GCN hidden layer, an LSTM hidden layer and a full-connected output layer, while the discriminative network takes the form of full-connected feedforward network.}
\label{fig}
\end{figure}

\subsection{The GCN Hidden Layer} 
We utilize GCN to model the local topology structure of each single graph snapshot in the dynamic network. GCN is an efficient variant of convolutional neural networks that can operate directly on graphs. Formally, assume that there are $N$ nodes with $M$-dimensional features (or attributes) in a static graph. The topological structure and node attributes can be respectively represented by an adjacency matrix ${\bf{A}} \in {\mathbb{R} ^{N \times N}}$ and a feature matrix ${\bf{Z}} \in {\mathbb{R} ^{N \times M}}$ (in which the $i$-th row of ${\bf{Z}}$ corresponds to the feature vector of node $i$). A typical GCN unit takes the feature matrix $\bf{Z}$ as the input and conducts the spectral graph convolution operation (according to $\bf{A}$) with localized first-order approximation on it. The final output is generated in the same way with a standard fully-connected layer. The overall operation of a specific GCN unit can be briefly defined as follows:
\begin{equation}
{\bf{X}} = GCN\left( {{\bf{Z}},{\bf{A}}} \right) = f\left( {{{{\bf{\hat D}}}^{ - {1 \mathord{\left/
 {\vphantom {1 2}} \right.
 \kern-\nulldelimiterspace} 2}}}{\bf{\hat A}}{{{\bf{\hat D}}}^{ - {1 \mathord{\left/
 {\vphantom {1 2}} \right.
 \kern-\nulldelimiterspace} 2}}}{\bf{ZW}}} \right),
\end{equation}
where ${{{\bf{\hat D}}}^{ - {1 \mathord{\left/{\vphantom {1 2}} \right.\kern-\nulldelimiterspace} 2}}}{\bf{\hat A}}{{{\bf{\hat D}}}^{ - {1 \mathord{\left/{\vphantom {1 2}} \right.\kern-\nulldelimiterspace} 2}}}$ is the approximated graph convolution filter with ${\bf{\hat A}} = {\bf{A}} + {{\bf{I}}_N}$ (${\bf{I}}_N$ is an $N$-dimensional identity matrix) and ${{{\bf{\hat D}}}_{ii}} = \sum\nolimits_{j = 1}^N {{{{\bf{\hat A}}}_{ij}}}$; $\bf{W}$ represents the weight matrix; $f\left( \cdot \right)$ is the activation function; $\bf{X}$ is the representation output given by the GCN unit.

For the temporal link prediction task that considers more than one static graph, the GCN-GAN model maintains a GCN unit $GCN\left( {{\bf{Z}},{{\bf{A}}_t}} \right) = {{\bf{X}}_t}$ for each graph snapshot input ${{\bf{A}}_t}$ ($t\in\left\{ {\tau - l, \cdots ,\tau } \right\}$). In our model, the feature matrix $\bf{Z}$ is set to be the noise input of the generative network, where the value of $\bf{Z}$ is generated according to a certain probability distribution $p$ (e.g. the uniform distribution). Based on the input of $\bf{Z}$ and ${\bf{A}}_{\tau - l}^\tau = \left\{{{{\bf{A}}_{\tau - l}}, \cdots ,{{\bf{A}}_\tau }}\right\}$, the GCN layer of the generative network outputs a sequence of representations notated as ${\bf{X}}_{\tau - l}^\tau = \left\{{{{\bf{X}}_{\tau - l}}, \cdots ,{{\bf{X}}_\tau }}\right\}$.

\subsection{The LSTM Hidden Layer} In the GCN-GAN model, the learned comprehensive network representations ${\bf{X}}_{\tau - l}^\tau$ are fed into an LSTM layer, which has a powerful capacity to learn the long-term dependencies of sequential data, to capture the evolving patterns of the weighted dynamic networks. The standard LSTM architecture can be described as an encapsulated cell with several multiplicative gate units. For a certain time step $t$, the LSTM cell takes current input vector ${{\bf{x}}_t}$ as well as the state vector of last time step ${{{\bf{h}}_{t-1}}}$ as the input, and then output the state vector in current time step ${{\bf{h}}_t}$:
\begin{gather}
{{\bf{i}}_t} = \sigma \left( {{\bf{W}}_x^i{{\bf{x}}_t} + {\bf{W}}_h^i{{\bf{h}}_{t - 1}} + {{\bf{b}}^i}} \right)\\
{{\bf{f}}_t} = \sigma \left( {{\bf{W}}_x^f{{\bf{x}}_t} + {\bf{W}}_h^f{{\bf{h}}_{t - 1}} + {{\bf{b}}^f}} \right)\\
{{\bf{o}}_t} = \sigma \left( {{\bf{W}}_x^o{{\bf{x}}_t} + {\bf{W}}_h^o{{\bf{h}}_{t - 1}} + {{\bf{b}}^o}} \right)\\
{{\bf{s}}_t} = {{\bf{f}}_t} \odot {{\bf{s}}_{t - 1}} + {{\bf{i}}_t} \odot \tilde{\bf{s}_t}\\
\tilde{\bf{s}_t} = \sigma \left( {{\bf{W}}_x^s{{\bf{x}}_t} + {\bf{W}}_h^s{{\bf{h}}_{t - 1}} + {{\bf{b}}^s}} \right)\\
{{\bf{h}}_t} = {{\bf{o}}_t} \odot \tanh \left( {{{\bf{s}}_t}} \right)
\end{gather}
where ${{\bf{i}}_t}$, ${{\bf{f}}_t}$, ${{\bf{o}}_t}$ and ${{\bf{s}}_t}$ represent the input gate, forget gate, output gate and memory cell, respectively; $\left\{ {{\bf{W}}_x ,{\bf{W}}_h ,{{\bf{b}}}} \right\}$ are the parameters of the corresponding unit;  $\sigma \left( \cdot \right)$ is the sigmoid activation function; $\odot$ denotes the element-wise multiplication.

Eventually, we treat the last hidden state ${{\bf{h}}_{\tau + 1}}$ as the distributed representation of the historical snapshots and feed it into a fully-connected layer to generate the prediction result ${{{\bf{\tilde A}}}_{\tau + 1}}$.

Due to the capacity of learning temporal information of sequential data, one can directly use the LSTM framework (with multiple inputs and single output) to tackle the temporal link prediction task. However, there remain some limitations for the prediction of the weighted dynamic networks. Specifically, to learn the temporal information of the dynamic network, LSTM is usually trained by using the Mean Square Error (MSE) loss function. However, the MSE loss cannot reflect the sparsity and wide-value-range of the link weights in the real network systems, which is empirically demonstrated in Section \uppercase \expandafter {\romannumeral5}.


\subsection{The Generative Adversarial Network}
To cope with the sparsity and wide-value-range problem of the dynamic network's edge weights, we utilize the GAN framework to enhance the generative capacity of LSTM.

Typically, GAN consists of a generative model $G$ and a discriminative model $D$ that compete in a minimax game with two players. First, $D$ tries to distinguish real data in the training set from the data generated by $G$. On the other hand, $G$ tries to fool $D$ and generate high-quality samples (data). Formally, such process can be described as follow (with two alternative optimization steps):
\begin{equation}
\mathop {\min }\limits_G \mathop {\max }\limits_D \left( \begin{array}{c}
{{\mathop{\rm E}\nolimits} _{x \sim {p_{data}}\left( x \right)}}\left[ {\log D\left( x \right)} \right] + \\
{{\mathop{\rm E}\nolimits} _{z \sim p\left( z \right)}}\left[ {\log \left( {1 - D\left( {G\left( z \right)} \right)} \right)} \right]
\end{array} \right),
\end{equation}
where $x$ is the input data from the training set, and $z$ represents the noise generated via a certain probability distribution $p\left(z \right)$ (e.g., the uniform distribution). 

Like the above standard GAN framework, our model also optimizes two neural networks (i.e., the generative network $G$ and the discriminative network $D$) with a minimax two-player game. In the model, $D$ tries to distinguish the real graph snapshot in the training data from the snapshot generated by $G$, while $G$ maximizes the probability of $D$ to make a mistake. Hopefully, this adversarial process can eventually adjust $G$ to generate plausible and high-quality weighted links. We further elaborate such two neural networks below.

\paragraph{The Discriminative Network $D$} As depicted in Fig. 1 (top side), we implement the discriminative model $D$ via a full-connected feedforward neural network with one hidden layer and one output layer. In the training process, $D$ alternatively takes $G$'s output ${{{\bf{\tilde A}}}_{\tau + 1}}$ or the ground-truth ${{\bf{A}}_{\tau + 1}}$ as the input. Since each input data of the full-connected neural network is usually represented as a vector (but not in the form of matrix), we reshape the matrix input (i.e., ${{{\bf{\tilde A}}}_{\tau + 1}}$ or ${{\bf{A}}_{\tau + 1}}$) into a corresponding row-wise long vector when feeding it into $D$. Moreover, as we adopt the Wasserstein GAN (WGAN) framework \cite{Arjovsky2017Towards, Arjovsky2017Wasserstein} to train the model (which is discussed later in this section), we set the output layer to be a linear layer, which directly generates the output without a non-linear activation function. Briefly, the details of the discriminative network $D$ can be formulated as follow:
\begin{equation}
D\left( {{\bf{A'}}} \right) = \left( {\sigma \left( {{\bf{a'W}}_h^D + {\bf{b}}_h^D} \right){\bf{W}}_o^D + {\bf{b}}_o^D} \right),
\end{equation}
where ${\bf{A'}} \in \left\{ {{{\bf{A}}_{\tau + 1}},{{{\bf{\tilde A}}}_{\tau + 1}}} \right\}$ with ${\bf{a}}'$ as the corresponding reshaped row-wise long vector; $\left\{ {{\bf{W}}_h^D,{\bf{b}}_h^D} \right\}$ and $\left\{ {{\bf{W}}_o^D,{\bf{b}}_o^D} \right\}$ are respectively the parameters of the hidden layer and the output layer; $\sigma \left( \cdot \right)$ represents the sigmoid activation function of the hidden layer.

Since the edge weights of a given network snapshot may have a large value range (e.g., [0, 2,000]), we normalize the element value of ${{\bf{A}}_{\tau + 1}}$ into the range of [0, 1] when selecting ${{\bf{A}}_{\tau + 1}}$ as the input of $D$. The original prediction result ${{{\bf{\tilde A}}}_{\tau  + 1}}$ given by $G$ is defined within the range [0, 1], so it can be directly utilized as the input of $D$.

\paragraph{The Generative Network $G$} As depicted in Fig. 1 (bottom side), the generative model $G$ consist of a GCN layer, an LSTM layer and a full-connected output layer.
The GCN layer takes the graph snapshots sequence ${\bf{A}}_{\tau - l}^\tau$ as well as the noise $\bf{Z}$ as the input, and outputs the representations sequence ${\bf{X}}_{\tau - l}^\tau$ which is later fed into the LSTM layer. Note that each adjacency matrix input ${{\bf{A}}_t}$ ($t \in \left\{ {\tau - l, \cdots ,\tau } \right\}$) should be normalized into the range of [0, 1] before being fed into the GCN layer. Moreover, we adopt sigmoid as the activation function of all the GCN units and let the noise input $\bf{Z}$ follow a uniform distribution within [0, 1] (notated as ${\bf{Z}} \sim {\mathop{\rm U}\nolimits} \left( {0,1} \right)$).

The LSTM layer takes the representations sequence ${\bf{X}}_{\tau - l}^\tau$ given by the GCN layer as the input, and outputs the hidden states ${\bf{h}}_{\tau - l}^\tau = \left\{ {{{\bf{h}}_{\tau - l}}, \cdots ,{{\bf{h}}_\tau}}\right\}$. Note that each matrix input ${{\bf{X}}_t}$ ($t \in \left\{ {\tau - l, \cdots ,\tau } \right\}$) should be reshaped into a row-wise vector ${{\bf{x}}_t}$ when being fed into the LSTM, since each LSTM cell treats the input data as a vector. Finally, the last hidden state ${{\bf{h}}_{\tau}}$ is fed into the output layer to generate the graph snapshot ${{{\bf{\tilde A}}}_{\tau + 1}}$ (with the corresponding row-wise vector form) of the next time slice. In particular, the elements of a certain generated result are within the range of [0, 1]. The final predicted snapshot ${{{\bf{\tilde A}}}_{\tau + 1}}$ with the correct value range of the dynamic network can be obtained by conducting the inverse process of normalization.

In the rest of this paper, we utilize the following simplified notation to represent the generative network $G$ (with noise $\bf{Z}$ and network snapshot sequence ${\bf{A}}_{\tau - l}^\tau$ as the input):
\begin{equation}
{{{\bf{\tilde A}}}_{\tau + 1}} = G\left( {{\bf{Z}},{\bf{A}}_{\tau - l}^\tau } \right).
\end{equation}

\subsection{Model Optimization}
Since the network's topology dynamically changes over time, the GCN-GAN model should constantly update its parameters to adapt to the network's evolution. Moreover, it's usually assumed that the network snapshot close to the next time slice $(\tau + 1)$ can be considered to have more similar characteristics to the ground-truth compared to those far from it \cite{Ma2018Graph}. Based on such reasonable assumption, we utilize the following optimization strategy. When it comes to a new time slice $\tau$, the model first conducts the training process by utilizing the previous network graph sequence ${\bf{A}}_{\tau - l - 1}^{\tau - 1}$ as the input and current snapshot ${{\bf{A}}_{\tau}}$ as the ground-truth. After training the model for current time slice $\tau$, we conduct the prediction process to generate the next graph snapshot ${{{\bf{\tilde A}}}_{\tau + 1}}$ with sequence ${\bf{A}}_{\tau - l}^\tau$ as the input. The details of the training and predicting process are elaborated in the rest of the section.

For the temporal link prediction task, directly utilizing the standard adversarial training process is inappropriate, because $G$ may generate a plausible network snapshot that can successfully fool $D$ but it is not consistent with the next graph snapshot. In fact, we expect that the prediction result should be as close as possible to the ground-truth ${{\bf{A}}_{\tau + 1}}$. In order to tackle such possible problem, we introduce another pre-training process for $G$ with the following loss function:
\begin{equation}
\begin{array}{l}
\mathop {\min }\limits_{{\theta _G}} h\left( {{\theta _G};{\bf{Z}},{\bf{A}}_{\tau  - l - 1}^{\tau  - 1},{{\bf{A}}_\tau }} \right) = \\
\left\| {{{\bf{A}}_\tau } - G\left( {{\bf{Z}},{\bf{A}}_{\tau  - l - 1}^{\tau  - 1}} \right)} \right\|_F^2 + \frac{\lambda }{2}\left\| {{\theta _G}} \right\|_2^2
\end{array},
\end{equation}
where ${\theta {}_G}$ represents the parameters of $G$ and $\lambda$ is the parameter to control the effect of the $L_2$-regularization term. In (12), $G$ tries to reconstruct current graph snapshot ${{{\bf{A}}_\tau }}$ given by the snapshot sequence ${{\bf{A}}_{\tau - l - 1}^{\tau - 1}}$ and noise $\bf{Z}$. Such process can help $G$ to fully capture the latest temporal information of the dynamic network, which is considered as the most similar characteristics to the real snapshot of ${{\bf{A}}_{\tau + 1}}$.

After the pre-training procedure, $G$ has the initial ability to generate the prediction result. The adversarial training process can be further developed to enhance $G$'s generative capacity to cope with the sparsity and wide-value-range problem of weighted dynamic networks. 
Especially, we utilize the Wasserstein GAN (WGAN) framework \cite{Arjovsky2017Towards, Arjovsky2017Wasserstein}, which has been proved to have a more reliable performance than the standard GAN, to achieve a relatively stable training process.

In such procedure, we first utilize the gradient descent method to update $D$'s parameters (notated as ${{\theta _D}}$) with $G$'s parameters fixed via the following loss function:
\begin{equation}
\begin{array}{l}
    \mathop {\min }\limits_{{\theta _D}} {h_D}\left( {{\theta _D};{\bf{Z}},{\bf{A}}_{\tau-l-1}^{\tau-1} ,{{\bf{A}}_{\tau}}} \right) = \\
    {\mathop{\rm E}\nolimits} \left[ {D\left( {{{\bf{A}}_{\tau}}} \right)} \right] - {\mathop{\rm E}\nolimits} \left[ {D\left( {G\left( {{\bf{Z}},{\bf{A}}_{\tau-l-1}^{\tau-1} } \right)} \right)} \right]
 \end{array}.
\end{equation}
After updating the parameters of $D$, their values should be further clipped into a pre-defined range of $[-c, c]$. Then, we update the parameters of $G$ (denoted as ${\theta_G}$) with $D$'s parameters fixed by using the following loss function: 
\begin{equation}
\begin{array}{l}
    \mathop {\min }\limits_{{\theta _G}} {h_G}\left( {{\theta _G};{\bf{Z}},{\bf{A}}_{\tau  - l - 1}^{\tau  - 1}} \right) = \\
    - {\rm{E}}\left[ {D\left( {G\left( {{\bf{Z}},{\bf{A}}_{\tau  - l - 1}^{\tau  - 1}} \right)} \right)} \right]
\end{array}.
\end{equation}
In the experiment, we adopted the RMSProp algorithm to alternatively update the parameters of ${{\theta _D}}$ and ${{\theta _G}}$ until converge.

After finishing the training process, $G$ can be utilized to generate the prediction result ${{{\bf{\tilde A}}}_{\tau + 1}}$ with ${\bf{A}}_{\tau - l}^\tau$ and $\bf{Z}$ as the input. Note that the original ${{{\bf{\tilde A}}}_{\tau + 1}}$'s elements are within the range of [0, 1], and another renormalization process is needed to recover its values to the real range of the network's edge weights. Furthermore, some tricks can be used to further refine the prediction result, which are formulated as follow:
\begin{gather}
{{{\bf{\tilde A}}}_{\tau  + 1}} \leftarrow {{\left( {{{{\bf{\tilde A}}}_{\tau  + 1}} + {\bf{\tilde A}}_{\tau  + 1}^T} \right)} \mathord{\left/ {\vphantom {{\left( {{{{\bf{\tilde A}}}_{\tau  + 1}} + {\bf{\tilde A}}_{\tau  + 1}^T} \right)} 2}} \right. \kern-\nulldelimiterspace} 2},\\
{\left( {{{{\bf{\tilde A}}}_{\tau + 1}}} \right)_{ii}} \leftarrow 0\mathop {}\limits_{} {\rm{for}}\mathop {}\limits_{} i \in \left\{ {1,2, \cdots ,\left| V \right|} \right\},\\
{\left( {{{{\bf{\tilde A}}}_{\tau + 1}}} \right)_{ij}} \leftarrow 0\mathop {}\limits_{} {\rm{if}}\mathop {}\limits_{} {\left( {{{{\bf{\tilde A}}}_{\tau + 1}}} \right)_{ij}} < \varepsilon.
\end{gather}
First, we use (15) to make ${{{\bf{\tilde A}}}_{\tau + 1}}$ symmetric, as we only consider the case of undirected networks. Then, by using (16), we set the diagonal elements of ${{{\bf{\tilde A}}}_{\tau + 1}}$ to be 0 to remove the effect of self-connected edges. Finally, the elements whose values are less than a small threshold $\varepsilon$ can be set to 0 to reflect the sparsity of edge weights.

We summarize the overall training and predicting procedures of the GCN-GAN model (when the network system comes to a new time slice $\tau$) in Table \uppercase \expandafter {\romannumeral1}.

\begin{table}[htbp]
\caption{The GCN-GAN Temporal Link Prediction Algorithm}
\begin{center}
\begin{tabular}{l}
\hline
\textbf{The GAN-CAN Algorithm} \\ \hline
\begin{tabular}[c]{@{}l@{}}
  \textbf{Input}: $\left\{ {{{\bf{A}}_{\tau - l - 1}}, \cdots ,{{\bf{A}}_{\tau - 1}},{{\bf{A}}_\tau }} \right\}$, $\left\{ {{\alpha _0},{\alpha _D},{\alpha _G}} \right\}$, $\left\{ {{n_0},n} \right\}$, $c$.\\
  
  \textbf{Output}: ${{{\bf{\tilde A}}}_{\tau + 1}}$.\\
  
  //$\left\{ {{{\bf{A}}_{\tau - l - 1}}, \cdots ,{{\bf{A}}_{\tau - 1}},{{\bf{A}}_\tau }} \right\}$: the network snapshot sequence\\
  //used to train the model and predict the next network snapshot;\\
  //$\left\{{{\alpha _0},{\alpha _D},{\alpha _G}} \right\}$: learning rate for pre-training and formal training;\\
  //$\left\{ {{n_0},n} \right\}$: number of iterations for pre-training and formal training;\\
  //$c$: clipping bound; ${{{\bf{\tilde A}}}_{\tau + 1}}$: prediction result.\\
  
  \hspace{10pt}//Train the GAN-GAN model\\
  \hspace{10pt}//with ${\bf{A}}_{\tau - l - 1}^{\tau - 1} = \left\{ {{{\bf{A}}_{\tau - l - 1}}, \cdots ,{{\bf{A}}_{\tau - 1}}} \right\}$ as the input\\
  \hspace{10pt}//and ${{\bf{A}}_\tau }$ as the ground-truth\\
  
  \hspace{10pt}normalize the values of $\left\{ {{{\bf{A}}_{\tau - l - 1}}, \cdots ,{{\bf{A}}_\tau }} \right\}$ into [0, 1]\\
  \hspace{10pt}\textbf{for} $i$ \textbf{from} $1$ \textbf{to} $n_0$ //Pre-train $G$\\
  \hspace{20pt}generate the noise input ${\bf{Z}} \sim {\mathop{\rm U}\nolimits} \left[ {0,1} \right]$\\
  \hspace{20pt}${\varsigma _{{\theta_G}}} \leftarrow {\nabla _{{\theta_G}}}h\left( {{\theta_G};{\bf{Z}},{\bf{A}}_{\tau - l - 1}^{\tau - 1},{{\bf{A}}_\tau }} \right)$\\
  \hspace{20pt}${\theta _G} \leftarrow {\theta _G} - {\alpha _0} \cdot {\mathop{\rm RMSProp}\nolimits} \left( {{\theta _G},{\varsigma _{{\theta _G}}}} \right)$\\
  
  \hspace{10pt}\textbf{for} $i$ \textbf{from} $1$ to $n$ //Formally train the model\\
  
  \hspace{20pt}generate the noise input ${\bf{Z}} \sim {\mathop{\rm U}\nolimits} \left[ {0,1} \right]$\\
  \hspace{20pt}${\varsigma_{{\theta _D}}} \leftarrow {\nabla_{{\theta _D}}}{h_D}\left( {{\theta _D};{\bf{Z}},{\bf{A}}_{\tau - l - 1}^{\tau - 1},{{\bf{A}}_\tau }} \right)$\\
  \hspace{20pt}${\theta _D} \leftarrow {\theta _D} - {\alpha _D} \cdot {\mathop{\rm RMSProp}\nolimits} \left( {{\theta _D},{\varsigma _{{\theta _D}}}} \right)$\\
  \hspace{20pt}clip the element value of  ${\theta _D}$ into $[-c, c]$\\
  
  \hspace{20pt}generate the noise input ${\bf{Z}} \sim {\mathop{\rm U}\nolimits} \left[ {0,1} \right]$\\
  \hspace{20pt}${\varsigma _{{\theta_G}}} \leftarrow {\nabla_{{\theta _G}}}{h_G}\left( {{\theta_G};{\bf{Z}},{\bf{A}}_{\tau - l - 1}^{\tau - 1}} \right)$\\
  \hspace{20pt}${\theta _G} \leftarrow {\theta _G} - {\alpha _G} \cdot {\mathop{\rm RMSProp}\nolimits} \left( {{\theta _G},{\varsigma _{{\theta _G}}}} \right)$\\
  
  \hspace{10pt}//Generate the prediction result ${{{\bf{\tilde A}}}_{\tau + 1}}$\\
  \hspace{10pt}//with ${\bf{A}}_{\tau - l}^\tau  = \left\{ {{{\bf{A}}_{\tau - l}}, \cdots ,{{\bf{A}}_\tau }} \right\}$ as the input\\
  \hspace{10pt}generate the noise input ${\bf{Z}} \sim {\mathop{\rm U}\nolimits} \left[ {0,1} \right]$\\
  \hspace{10pt}${{{\bf{\tilde A}}}_{\tau + 1}} \leftarrow G\left( {{\bf{Z}},{\bf{A}}_{\tau - l}^\tau } \right)$\\
  \hspace{10pt}renormalize ${{{\bf{\tilde A}}}_{\tau + 1}}$ into the real value range\\
  \hspace{10pt}refine ${{{\bf{\tilde A}}}_{\tau + 1}}$ via (15), (16) and (17)
  
\end{tabular} \\ 
\hline
\end{tabular}
\end{center}
\end{table}

\section{Experimental Evaluation}

\subsection{Datasets}
In order to evaluate the effectiveness of our model, we conduct experiments on three real datasets and one simulation dataset of different network systems. The detailed statistics of these four datasets are presented in Table \uppercase\expandafter{\romannumeral2}, where $N$, $T$ and $Max$ denote the number of nodes, the number of time slices and the maximum edge weight value for a certain dataset after necessary pre-processing.
\begin{table}[htbp]
\caption{Details of the datasets}
\begin{center}
\begin{tabular}{l|llll}
\hline
\textbf{Datasets} & \textit{\textbf{N}} & \textit{\textbf{T}} & \textit{\textbf{Max}} & \textbf{Description}\\ \hline
\textit{UCSB}  & 38 & 1,000 & 2,000 & Wireless mesh net link quality\\
\textit{KAIST}  & 92 & 500 & 250 & Human mobility position data\\
\textit{BJ-Taxi}  & 256 & 500 & 2,000 & Vehicle mobility position data\\
\textit{NumFabric}  & 128 & 350 & 20,000 & Simulation data center flow\\ \hline
\end{tabular}
\end{center}
\end{table}

\textit{UCSB}\footnote{https://crawdad.org/ucsb/meshnet/20070201/} \cite{Ramachandran2007Routing} is a link quality dataset of a wireless mesh network; \textit{NumFabric}\footnote{The data can be generated by running the simulation code released in https://knagaraj@bitbucket.org/knagaraj/numfabric.git} \cite{Nagaraj2016NUMFabric} is a simulation flow dataset of data center; \textit{KAIST}\footnote{https://crawdad.org/ncsu/mobilitymodels/20090723/} \cite{Lee2009SLAW} and \textit{BJ-Taxi}\footnote{https://www.microsoft.com/en-us/research/publication/t-drive-driving-directions-based-on-taxi-trajectories/} \cite{Yuan2011Driving} are the position datasets of a human mobility network and a vehicle mobility network, respectively. (Both \textit{KAIST} and \textit{BJ-Taxi} are the subsets of the original datasets.) For each dataset, we pre-process the dynamic network into multiple successive graph snapshots. 

With regard to \textit{UCSB} and \textit{NumFabric}, the hosts in the network systems can be described as the nodes in the dynamic networks. Moreover, the link quality or flow in a certain time slice can be directly represented as the link weight between the corresponding pair of hosts in the specific snapshot.

For the position datasets \textit{KAIST} and \textit{BJ-Taxi}, we treated each user (person or vehicle) as a node in the abstracted dynamic network and calculated the distance between each pair of nodes for all the time slices. Particularly, we constructed a distance matrix ${{\bf{D}}_t}$ for a specific time slice $t$, with ${\left( {{{\bf{D}}_t}} \right)_{ij}} = {\left( {{{\bf{D}}_t}} \right)_{ji}}$ representing the distance between node $i$ and $j$. In real network systems of human mobility and vehicle mobility, the pair of users who are close to each other should be given more interest or attention compared to those with relatively large distance. Hence, we constructed an abstracted weighed network based on the distance matrix ${{{\bf{D}}_t}}$, in which the link weights were inversely proportional to the corresponding distance. When pre-processed the \textit{KAIST} and \textit{BJ-Taxi} datasets, we utilized the following formulation to construct the weighted adjacency matrix ${{\bf{A}}_t}$ of the graph snapshot in time slice $t$:
\begin{equation}
{\left( {{{\bf{A}}_t}} \right)_{ij}} = \left\{ {\begin{array}{*{20}{c}}
{0,\begin{array}{*{20}{c}}
{\begin{array}{*{20}{c}}
{}&{}
\end{array}}&{}
\end{array}{\rm{if }}\mathop {}\limits_{} {{\left( {{{\bf{D}}_t}} \right)}_{ij}} \ge \delta }\\
{\delta  - {{\left( {{{\bf{D}}_t}} \right)}_{ij}},{\rm{if }}\mathop {}\limits_{} {{\left( {{{\bf{D}}_t}} \right)}_{ij}} < \delta }
\end{array}} \right.,
\end{equation}
where we set the link weight to be 0 if the corresponding distance was larger than a pre-set threshold $\delta$. Particularly, $\delta$ is the maximum edge weight for a certain network (with $\delta = 250$ for \textit{KASIT} and $\delta = 2,000$ for \textit{BJ-Taxi} in our experiments).

Furthermore, we also statistically analyze the sparsity and distribution of the link weights for the above four networks. 

\textbf{The sparsity of link weights.} According to our observation, most of the graph snapshots are relatively sparse, which means there is a non-ignorable portion of zeros in the adjacency matrix of a certain time slice. For example, the average portions of zero elements in all the adjacency matrices for \textit{UCSB}, \textit{KASIT}, \textit{BJ-Taxi} and \textit{NumFabric} are \textbf{0.52}, \textbf{0.92}, \textbf{0.94} and \textbf{0.50}, respectively. Although the sparsity degree of different network systems may be significantly different, such statistical result can still reveal the fact that there are usually non-ignorable pairs of entities in the network system that may not have the defined relation (e.g., the relation of data transmission in the data center).

For the weighted temporal link prediction task, ${\left( {{{\bf{A}}_t}} \right)_{ij}} = 0$ means there is no edge between node $i$ and $j$. On the other hand, ${\left( {{{\bf{A}}_t}} \right)_{ij}}$ with small value means there is still an edge between this pair of nodes but the weight is small. Such two circumstances are entirely different, but to distinguish them remains a challenging problem for most of the conventional temporal link prediction methods.

In fact, the mistake that a prediction model fails to distinguish small edge weights and zero values is relatively serious for most network systems. It may mistakenly direct the system to (\romannumeral1) pre-allocate the key resources for the nonexistence links or (\romannumeral2) not allocate resources for the existent links, resulting in more waste of system overhead than the normal prediction error of the existent links' weights.

\textbf{The distribution of edge weights.} The wide value range of link weights (e.g., from 0 to 20,000) is another significant property we have observed. To ensure the learning ability of the model, most link prediction approaches normalize the link weights input into a certain small range (e.g. $[0, 1]$). However, when recovering the value of the predicted snapshot, the very slight errors (between 0 and 1) may still result in large errors in the final results with the mean square error evaluation metric. 

In this study, we also investigate the distribution of edge weights for the four datasets. The statistic of $UCSB$ is illustrated in Fig. 2 as an example, in which $w$ (the horizontal axis) represents the possible edge weight value of the dataset, while $c(w)$ represents the number of edges with value $w$. Note that we use the logarithm of $c(w)$ as the vertical axis.
\begin{figure}[htbp]
\centerline
{\includegraphics[width=0.50\textwidth, trim=10 0 25 10, clip]{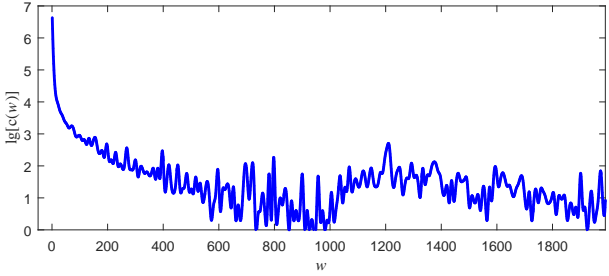}}
\caption{The distribution of the edge weights of \textit{UCSB}, in which the horizontal axis ($w$) represents the possible edge weight value while the logarithm of $c(w)$ is used as the vertical axis. $c(w)$ is the number of edges with value $w$.}
\label{fig}
\end{figure}

As shown in Fig. 2, a large proportion of edges have small weights in the network. However, the conventional Mean Square Error (MSE) metric used to train and evaluate the temporal link prediction model is only sensitive to edges with large weights. It cannot reflect the dynamic changing of the majority of the edges with small weights, making the temporal link prediction of weighted networks a challenging problem.

\subsection{Evaluation Metrics}
To quantitatively evaluate our temporal link prediction model, we follow previous work \cite{Ma2017Nonnegative, Ma2018Graph} to use the Mean Square Error (MSE) scores for comparison, which is defined as:
\begin{equation}
MSE = {{\left\| {{{\bf{A}}_{\tau {\rm{ + }}1}} - {{{\bf{\tilde A}}}_{\tau + 1}}} \right\|_F^2} \mathord{\left/
 {\vphantom {{\left\| {{{\bf{A}}_{\tau {\rm{ + }}1}} - {{{\bf{\tilde A}}}_{\tau  + 1}}} \right\|_F^2} {\left( {\left| V \right| \times \left| V \right|} \right)}}} \right.
 \kern-\nulldelimiterspace} {\left( {\left| V \right| \times \left| V \right|} \right)}}.
\end{equation}
Moreover, in order to further evaluate the capacity of our model to cope with the sparsity and wide-value-range problem discussed above, we introduce two additional metrics (i.e., the edge-wise KL-divergence and the mismatch rate).

\paragraph{The Edge-wise KL-Divergence} For some dynamic networks, the edge weights of a snapshot may have a wide value range (e.g., [0, 2,000]), in which the majority of the edges are with relatively small weights. Nevertheless, the MSE score may be sensitive to the large edge weights only and suffers from distinguishing the difference of magnitude that is important for small weights. For instance, the magnitude difference between 2 and 1 should be much larger than the difference between 2,000 and 1,990, even though the latter case results in larger MSE error. To alleviate the mentioned issue, we introduce the edge-wise KL-divergence to further consider the magnitude difference of link weights.


Firstly, we derive two auxiliary matrices  $\bf{P}$ and $\bf{Q}$ to represent the normalized values of the ground-truth graph snapshot ${{{\bf{A}}_{\tau + 1}}}$ and the prediction result ${{{{\bf{\tilde A}}}_{\tau + 1}}}$, respectively. We formulate $\bf{P}$ and $\bf{Q}$ as follows: 
\begin{equation}
{{\bf{P}}_{ij}} = \frac{{{{\left( {{{\bf{A}}_{\tau  + 1}}} \right)}_{ij}}}}{{\sum\nolimits_{i,j = 1}^N {{{\left( {{{\bf{A}}_{\tau  + 1}}} \right)}_{ij}}} }},
{\rm{ }}{{\bf{Q}}_{ij}} = \frac{{{{\left( {{{{\bf{\tilde A}}}_{\tau  + 1}}} \right)}_{ij}}}}{{\sum\nolimits_{i,j = 1}^N {{{\left( {{{{\bf{\tilde A}}}_{\tau  + 1}}} \right)}_{ij}}} }}.
\end{equation}

Then, the edge-wise KL-divergence is defined as:
\begin{equation}
KL\left( {{\bf{P}}\left\| {\bf{Q}} \right.} \right) = \sum\nolimits_{i,j = 1}^N {f\left( {{{\bf{P}}_{ij}},{{\bf{Q}}_{ij}}} \right)},
\end{equation}
where $f\left( {{{\bf{P}}_{ij}},{{\bf{Q}}_{ij}}} \right) = {{\bf{P}}_{ij}}\log \left( {{{{{\bf{P}}_{ij}}} \mathord{\left/{\vphantom {{{{\bf{P}}_{ij}}} {{{\bf{Q}}_{ij}}}}} \right. \kern-\nulldelimiterspace} {{{\bf{Q}}_{ij}}}}} \right)$ (with the standard form of the KL-divergence) if ${{{\bf{P}}_{ij}}>0}$ and ${{{\bf{Q}}_{ij}}>0}$; otherwise $f\left( {{{\bf{P}}_{ij}},{{\bf{Q}}_{ij}}} \right) = 0$. Note that  when ${{{\bf{P}}_{ij}}=0}$ or ${{{\bf{Q}}_{ij}}=0}$, we simply set their KL-divergence to 0 since the zero value may cause the NaN or Inf exception, and we consider such special cases in the definition of the mismatch rate below.


\paragraph{The Mismatch Rate} According to our observation on the datasets, the sparsity issue of the edge weights in the weighted dynamic network is also significant, which need to be specially discussed. We consider the following two cases:

(a) ${\left( {{{\bf{A}}_{\tau  + 1}}} \right)_{ij}} = 0$ but ${\left( {{{{\bf{\tilde A}}}_{\tau  + 1}}} \right)_{ij}} > 0$;

(b) ${\left( {{{\bf{A}}_{\tau  + 1}}} \right)_{ij}} > 0$ but ${\left( {{{{\bf{\tilde A}}}_{\tau  + 1}}} \right)_{ij}} = 0$.\\
Particularly, such two cases mean the prediction results improperly determine the existence of the edge $(i, j)$, which should be considered as serious mistakes for the temporal link prediction of weighted dynamic networks. Hence, we use the mismatch rate, which represents the proportion of such mismatched edges in a certain graph snapshot, as an additional evaluation metric.

\subsection{Performance Evaluation}
We evaluate the effectiveness of our model by comparing it with six baseline methods on the four datasets, including \textit{ED} \cite{Ma2017Nonnegative}, \textit{SVD} \cite{Ma2017Nonnegative}, \textit{NMF} \cite{Lee1999Learning}, \textit{GrNMF} \cite{Ma2018Graph}, \textit{AM-NMF} \cite{Kai2018Adaptive} and \textit{LSTM} \cite{Gers2014Learning}. Among the baselines, \textit{ED}, \textit{SVD} and \textit{NMF} are conventional approaches based on the collapsed network\cite{Liben2007The, Sharan2008Temporal}, while \textit{GrNMF} and \textit{AM-NMF} are state-of-the-art matrix factorization-based methods without collapsing the dynamic network. \textit{LSTM} represents the non-linear model that directly utilize LSTM \cite{Gers2014Learning} to perform the temporal link prediction.

In the experiment, we uniformly set $l=10$ for all the methods to be evaluated. For the matrix factorization-based methods (i.e., \textit{ED}, \textit{SVD}, \textit{NMF}, \textit{GrNMF} and \textit{AM-NMF}), we set the hidden space size to be 16, 64, 128 and 64 for \textit{UCSB}, \textit{KAIST}, \textit{BJ-Taxi} and \textit{NumFabric}, respectively. For \textit{ED}, \textit{SVD} as well as \textit{NMF}, we select the parameters with the best performance, while we utilize the recommended parameter settings for \textit{GrNMF} and \textit{AM-NMF}.

With regard to the non-linear methods (i.e., \textit{LSTM} and \textit{GCN-GAN}), we use the Xavier approach \cite{Glorot2010Understanding} to initialize the parameters. For a dataset with $N$ nodes, we set the noise input ${\bf{Z}}$ and the adjacency matrix input ${{\bf{A}}_t}$ of the GCN-GAN model to have the same size of $N \times N$. Note that the input of a GCN unit is in the form of matrix, and its output should be reshaped into a row-wise vector before being fed into the LSTM layer. The layer configurations of the four datasets with the format of ${m_i}$-${m_h}$-${m_o}$ are illustrated in Table \uppercase\expandafter{\romannumeral3}, where ${m_i}$ is the size of the input in each time step, ${m_h}$ represents the hidden size of LSTM and ${m_o}$ is the size of the output. Especially, we use $N \times N$ to represent the matrix input (with the size of $N \times N$) and utilize ${N^2}$ to represent the size of a (row-wise) long vector which can be reshaped into an $N \times N$ matrix.

\begin{table}[htbp]
\caption{The Layer Configurations of GCN-GAN and LSTM.}
\begin{center}
\begin{tabular}{l|l|l|c}
\hline
\multicolumn{1}{c|}{\multirow{2}{*}{\textbf{Datasets}}} & \multicolumn{2}{c|}{\textbf{GCN-GAN}} & \multirow{2}{*}{\textbf{LSTM}} \\ \cline{2-3}
\multicolumn{1}{c|}{} & \multicolumn{1}{c|}{\textit{\textbf{G}}} & \multicolumn{1}{c|}{\textit{\textbf{D}}} &  \\ \hline
\textit{\textbf{UCSB}} & (38$\times$38)-38-$38^2$ & $38^2$-512-1 & $38^2$-128-$38^2$ \\
\textit{\textbf{KAIST}} & (92$\times$92)-92-$92^2$ & $92^2$-512-1 & $92^2$-128-$92^2$ \\
\textit{\textbf{BJ-Taxi}} & (256$\times$256)-256-$256^2$ & $256^2$-1024-1 & $256^2$-512-$256^2$ \\
\textit{\textbf{NumFabric}} & (128$\times$128)-128-$128^2$ & $128^2$-1024-1 & $128^2$-512-$128^2$ \\ \hline
\end{tabular}
\end{center}
\end{table}

The parameter settings of the four datasets are shown in Table \uppercase\expandafter{\romannumeral4}, where $\lambda$ is the parameter controls the effect of the $L_2$-regularization term in (12); $\varepsilon$ is the threshold used to refine the prediction result in (17); ${\alpha_0}$ is the learning rate of the pre-training process; ${\alpha_D}$ and ${\alpha_G}$ are the learning rate for training $D$ and $G$; $c$ is the clipping bound for $D$'s parameters ${\theta_D}$.

\begin{table}[htbp]
\caption{The Parameter Settings of GCN-GAN.}
\begin{center}
\begin{tabular}{l|llllll}
\hline
\multicolumn{1}{c|}{\multirow{2}{*}{\textbf{Datasets}}} & \multicolumn{6}{c}{\textbf{Parameters}} \\ \cline{2-7} 
\multicolumn{1}{c|}{} & $\lambda$ & $\varepsilon$ & ${\alpha_0}$ & ${\alpha_D}$ & ${\alpha_G}$ & $c$ \\ \hline
\textit{\textbf{UCSB}} & 0 & 0.01 & 0.005 & 0.001 & 0.001 & \multirow{4}{*}{0.01} \\
\textit{\textbf{KAIST}} & 1e-5 & 0.01 & 0.01 & 0.0005 & 0.0005 &  \\
\textit{\textbf{BJ-Taxi}} & 1e-5 & 0.01 & 0.005 & 0.001 & 0.001 &  \\
\textit{\textbf{NumFabric}} & 0 & 0.5 & 0.001 & 0.001 & 0.001 &  \\ \hline
\end{tabular}
\end{center}
\end{table}

For each dataset, we use the first $(l+2)$ graph snapshots as the original training data to conduct the initial training process. Then, we continuously slide the window with size $(l+1)$ for the rest snapshots, in which we alternatively conduct the training and predicting process. For each evaluation metric, we record the average performance value of all the predicted snapshots. The evaluation results in terms of MSE, edge-wise KL-divergence and mismatch rate are shown in Table \uppercase\expandafter{\romannumeral5}, Table \uppercase\expandafter{\romannumeral6} and Table \uppercase\expandafter{\romannumeral7} respectively, where the best performance value is in \textbf{bold} and the second-best is with \underline{underline}.

\begin{table}[htbp]
\caption{Performance Evaluation Results in terms of MSE.}
\begin{center}
\begin{tabular}{l|llll}
\hline 
\multicolumn{1}{c|}{\multirow{2}{*}{\textbf{Methods}}} & \multicolumn{4}{c}{\textbf{Datasets}} \\ \cline{2-5} 
\multicolumn{1}{c|}{} & \textit{\textbf{UCSB}} & \textit{\textbf{KAIST}} & \textit{\textbf{BJ-Taxi}} & \textit{\textbf{NumFabric}} \\ \hline
\textit{ED} & 8.1118 & 0.5067 & 0.3134 & 33.6076 \\
\textit{SVD} & 5.4185 & 0.2043 & 0.2789 & 3.9849 \\
\textit{NMF} & 5.8586 & 0.2683 & 0.2825 & 4.0368 \\
\textit{GrNMF} & 5.6767 & 0.1381 & 0.2342 & 2.6210 \\
\textit{AM-NMF} & 5.6716 & 0.1380 & 0.2343 & 2.6364 \\
\textit{LSTM} & {\underline{5.2518}} & {\underline{0.1196}} & {\textbf{0.2283}} & {\underline{0.6867}} \\
\textit{\textbf{GCN-GAN}} & \textbf{5.1154} & \textbf{0.1189} & {\underline{0.2284}} & {\textbf{0.6570}} \\ \hline
\end{tabular}
\end{center}
\end{table}

\begin{table}[htbp]
\caption{Performance Evaluation Results in terms of Edge-wise KL-divergence.}
\begin{center}
\begin{tabular}{l|llll}
\hline
\multicolumn{1}{c|}{\multirow{2}{*}{\textbf{Method}}} & \multicolumn{4}{c}{\textbf{Datasets}} \\ \cline{2-5} 
\multicolumn{1}{c|}{} & \textit{\textbf{UCSB}} & \textit{\textbf{KAIST}} & \textit{\textbf{BJ-Taxi}} & \textit{\textbf{NumFabric}} \\ \hline
\textit{ED} & 0.5960 & 0.3619 & 1.0541 & 0.0681 \\
\textit{SVD} & {\underline{0.4726}} & 0.1351 & 1.0085 & 0.0085 \\
\textit{NMF} & 0.6980 & 0.1021 & 1.2519 & 0.0094 \\
\textit{GrNMF} & 0.7696 & 0.0653 & {\underline{0.4674}} & 0.0073 \\
\textit{AM-NMF} & 0.7981 & 0.0649 & 0.4752 & 0.0073 \\
\textit{LSTM} & 0.6120 & {\underline{0.0578}} & 1.3892 & {\underline{0.0012}} \\
\textit{\textbf{GCN-GAN}} & \textbf{0.3247} & \textbf{0.0262} & {\textbf{0.2831}} & {\textbf{0.0009}} \\ \hline
\end{tabular}
\end{center}
\end{table}

\begin{table}[htbp]
\caption{Performance Evaluation Results in terms of Mismatch Rate.}
\begin{center}
\begin{tabular}{l|llll}
\hline
\multicolumn{1}{c|}{\multirow{2}{*}{\textbf{Method}}} & \multicolumn{4}{c}{\textbf{Datasets}} \\ \cline{2-5} 
\multicolumn{1}{c|}{} & \textit{\textbf{UCSB}} & \textit{\textbf{KAIST}} & \textit{\textbf{BJ-Taxi}} & \textit{\textbf{NumFabric}} \\ \hline
\textit{ED} & 0.2227 & 0.1907 & 0.1923 & 0.4806 \\
\textit{SVD} & 0.1801 & 0.2466 & 0.2425 & 0.0003 \\
\textit{NMF} & 0.1651 & 0.1638 & 0.1932 & \underline{0.0001} \\
\textit{GrNMF} & {\underline{0.1264}} & {\underline{0.0912}} & {\underline{0.0258}} & \underline{0.0001} \\
\textit{AM-NMF} & 0.1305 & 0.0918 & 0.0283 & \underline{0.0001} \\
\textit{LSTM} & 0.2905 & 0.1175 & 0.9118 & 0.4923 \\
\textit{\textbf{GCN-GAN}} & \textbf{0.0133} & \textbf{0.0122} & \textbf{0.0173} & \textbf{3e-5} \\ \hline
\end{tabular}
\end{center}
\end{table}

Note that the precision differences of the evaluation metrics among the comparative methods in different datasets are different, since different network systems may have diverse edge weight ranges and temporal characteristics. In Table \uppercase\expandafter{\romannumeral5}, the non-linear models (i.e., \textit{LSTM} and \textit{GCN-GAN}) achieve much better MSE scores than the linear approaches, indicating the powerful feature learning ability of the non-linear deep neural network. Particularly, our method has the best MSE scores on three datasets (i.e., \textit{UCSB}, \textit{KAIST} and \textit{NumFabric}) and still obtains the competitive second-best MSE with the best performer (i.e., \textit{LSTM}) in \textit{BJ-Taxi}. For edge-wise KL-divergence and mismatch rate, the proposed method outperforms all the baselines on the four datasets, which further verifies that GCN-GAN is able to alleviate the sparsity and wide-value-range problem of weighted dynamic networks.

\subsection{Case Study}
We use an exemplary case, which is selected from the prediction results of \textit{UCSB}, to demonstrate our model's capacity to generate high-quality weighted links. For a certain graph snapshot, we compare the adjacency matrices generated by different methods with the ground-truth. Heat maps are used to visualize the results. Typically, in an adjacency matrix, the zero elements and those with small values have entirely different physical meanings, thus we specially set all the zero values in the matrix to $-200$ to emphasize such difference.

The results are reported in Fig. 3, in which subfigure (a) presents the ground-truth while subfigures (b) and (c) show the prediction results of \textit{GCN-GAN} and \textit{LSTM}, respectively. In the heat map, black represents zero value of the adjacency matrix (note that we set values of all the zero elements to be $-200$). Moreover, the color depth indicates the weights of the edges, in which the color close to dark red indicates relatively small edge weight while that close to white means large value.


As illustrated in Fig. 3, both \textit{GCN-GAN} and \textit{LSTM} can fit the large edge weight (with color close to white) well. However, the \textit{LSTM} fails to distinguish the zero values (with color of black) and small edge weights (with color close to dark red). On the other hand, our \textit{GCN-GAN} model can effectively cope with such challenging problem, reflecting the sparsity of edge weights of the network snapshot.



\begin{figure}
\centering
\begin{minipage}{0.45\linewidth}
\subfigure[Ground-truth]
{\includegraphics[width=\textwidth, trim=0 0 50 0, clip]{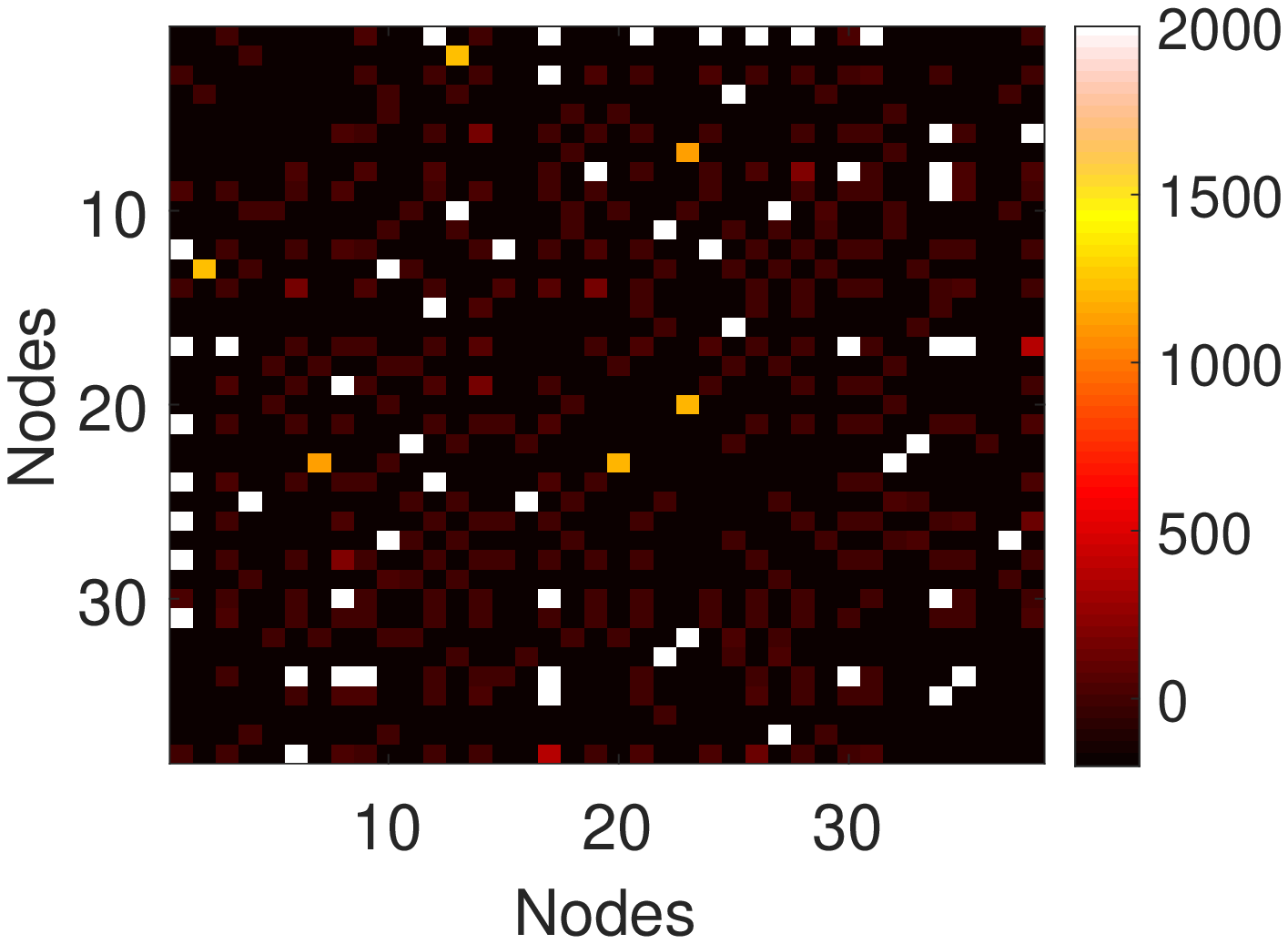}}
\end{minipage}
\begin{minipage}{0.45\linewidth}
\subfigure[\textit{GCN-GAN}]
{\includegraphics[width=\textwidth, trim=0 0 50 0, clip]{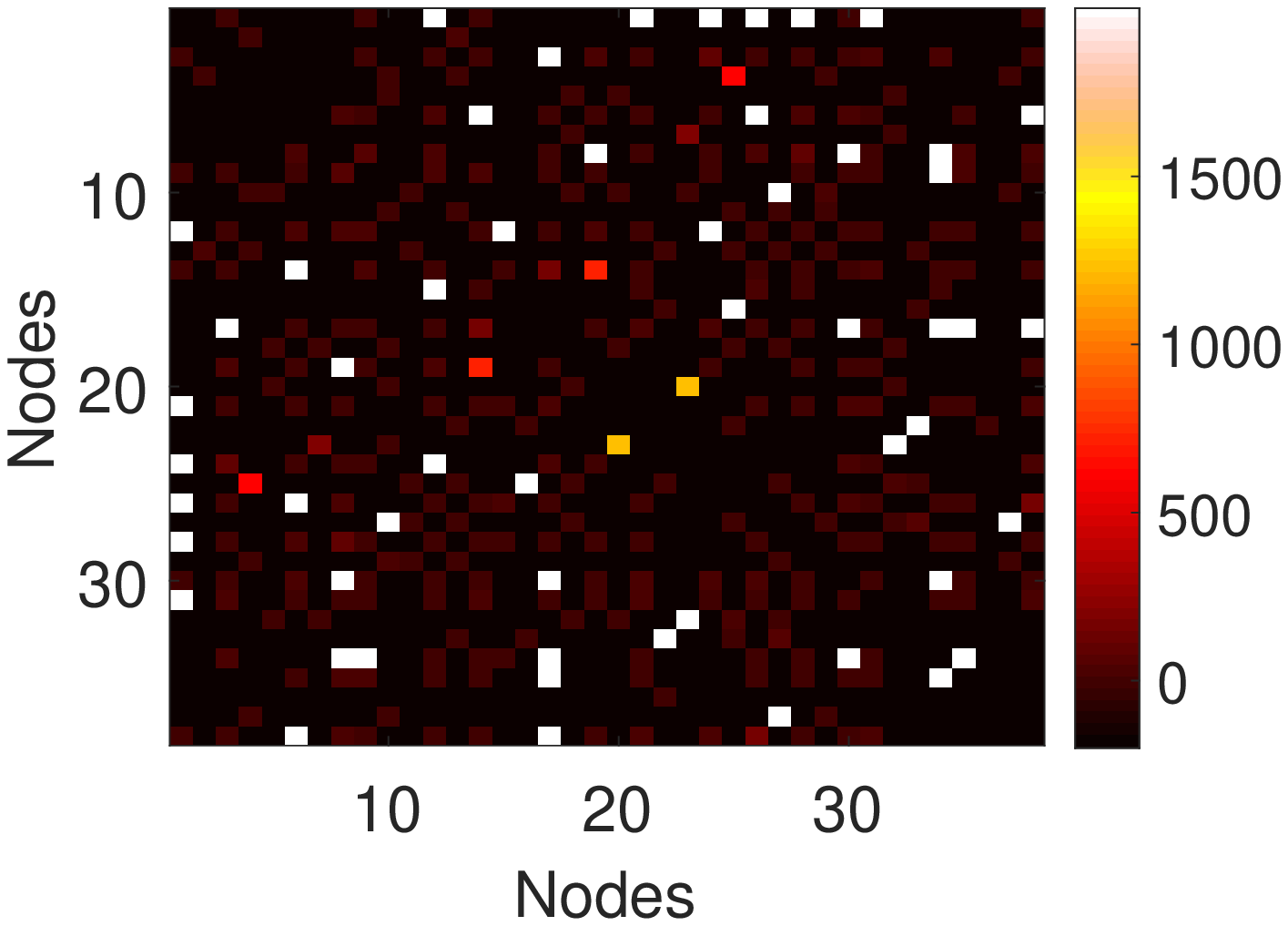}}
\end{minipage}

\begin{minipage}{0.45\linewidth}
\subfigure[\textit{LSTM}]
{\includegraphics[width=\textwidth, trim=0 0 50 0, clip]{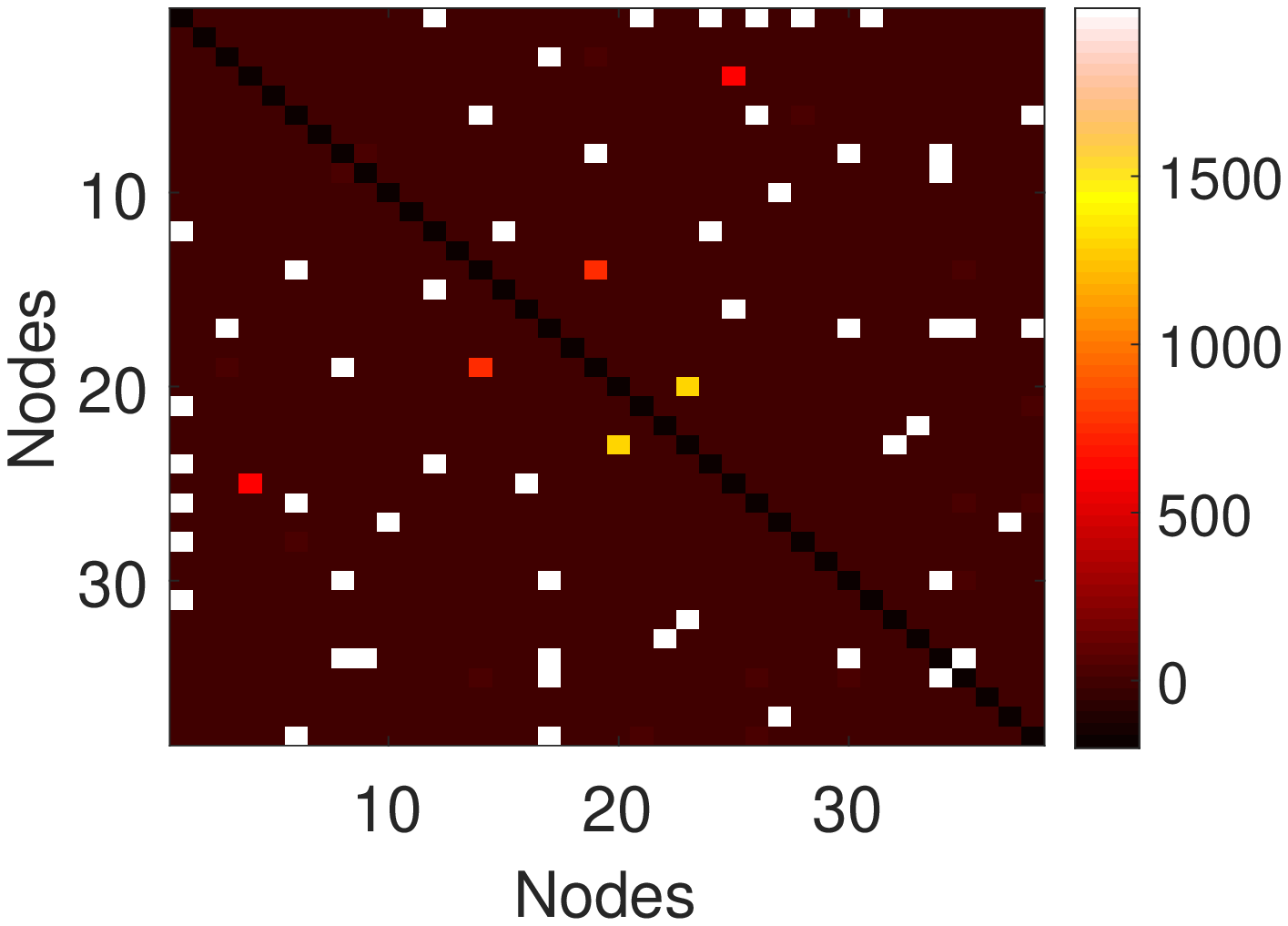}}
\end{minipage}
\begin{minipage}[c]{0.45\linewidth}
\caption{Adjacency matrices of (a) the ground-truth, (b) the prediction result of \textit{GCN-GAN} and (c) the prediction result of \textit{LSTM} corresponding to one graph snapshot in \textit{UCSB}.}
\end{minipage}

\end{figure}


\section{Conclusion}
In this paper, we proposed a novel temporal link prediction model GCN-GAN to generally tackle the dynamics prediction problem in network systems (e.g., the prediction of mobility, traffic and topology). Our model can effectively deal with challenging prediction task of weighted dynamic networks, because it combined the strengths of the deep neural networks (i.e., GCN and LSTM) in learning the comprehensive distributed representations of networks as well as GAN in generating high-quality weighted links. In addition, we applied the proposed model to four datasets of different network systems and specially analyzed the sparsity as well as the wide-value-range properties of the edge weights in real-life network systems. The performance evaluation results demonstrated that the proposed model outperformed other six competitors while having the powerful capacity to tackle the sparsity and wide-value-range problem of weighted dynamic networks.


In our future work, we will consider the concrete deployment scenario of real network systems and use the measures of communication networks to evaluate the performance improvement for different systems. More importantly, how to cope with the challenging temporal link prediction problem with an unfixed node set is also our next research focus.



\section{Acknowledgment}
This work has been funded by Natural Science Foundation of Guangdong (No.2018A030313017), Huawei Innovation Research Program (YBN2017125201) and Shenzhen Key Fundamental Project (JCYJ20170412151008290 and JCYJ20170412150946024) .

\bibliographystyle{IEEEtran}
\bibliography{GCN-GAN-Ref}

\end{document}